\documentstyle[aps,floats,epsf,twocolumn]{revtex}
\newcommand{\be}{\begin{equation}}
\newcommand{\ee}{\end{equation}}
\newcommand{\ba}{\begin{eqnarray}}
\newcommand{\ea}{\end{eqnarray}}
\newcommand{\nn}{\nonumber}

\begin{document} 

\title{Effective ``Penetration Depth" in the Vortex State of a
d-wave Superconductor}

\author{M. H. S. Amin$^{1\dagger}$
M. Franz$^2$ and Ian Affleck$^{3,4}$}
\address{$^1$ Department of Physics, Simon Fraser University,
Burnaby, BC, V5A 1S6, Canada, \\
$^2$Department of Physics and Astronomy, Johns Hopkins University,
Baltimore, MD 21218, USA,\\
$^3$Department of Physics and Astronomy
and Canadian Institute for Advanced
Research,\\ University of British Columbia, Vancouver, BC, V6T 1Z1, Canada\\
$^4$ Institute for Theoretical Physics, University of California,
Santa Barbara, CA93106-4030, USA \\
{\rm(\today)}}\maketitle

\begin{abstract}

The temperature and field dependence of the effective magnetic penetration depth 
($\lambda_{\rm eff}$) in the vortex state of a d-wave superconductor, as 
measured by muon spin rotation ($\mu$SR) experiments,
is calculated using a nonlocal London model. We show that at temperatures
below $T^* \propto \sqrt{B}$, 
the linear $T$-dependence of $\lambda_{\rm eff}^{-2}$ crosses over
to a $T^3$-dependence. This could provide an explanation for the 
low temperature flattening of the $\lambda_{\rm eff}^{-2}$ curve 
observed in a recent $\mu$SR experiment. 

\end{abstract}

\vspace{5pt}

Recent experiments on quasiparticle response in the vortex state of the  
cuprate superconductors 
have indicated quite unexpected behavior. Krishana {\it et al.}\ discovered
an anomalous plateau in the longitudinal thermal conductivity
$\kappa_{xx}$ of ${\rm Bi_2Sr_2CaCu_2O_{8+\delta}}$ (BSCCO)
\cite{krishana1} and ${\rm YBa_2Cu_3O_{7-\delta}}$ (YBCO)  \cite{ong} at
high magnetic fields which they attributed to the opening of a second gap
of $d_{xy}$ symmetry. Scanning tunneling spectroscopy (STS) on YBCO
compounds \cite{sts} lent support to this scenario by suggesting the
existence of localized quasiparticles in the vortex core regions which is
possible only for a gapped excitation spectrum \cite{ft}. 
 Subsequent measurements of Aubin {\it et al}.\
\cite{aubin1} discovered hysteretic behavior in the thermal conductivity
which is a signature of the influence of the vortex lattice and impurities
on $\kappa_{xx}$. Furthermore, at temperatures below 1K they detected an
increase of $\kappa_{xx}$ with the magnetic field instead of decrease
\cite{aubin2}.  This is actually in agreement with the existence of nodes
in the superconducting gap and in contradiction with $d_{x^2-y^2}+id_{xy}$
symmetry scenario.  Theoretically, the  possibility
of a field induced $d_{xy}$ gap was discussed by Laughlin and others
\cite{laughlin}, and attempts have been made to explain the observed
strange behavior of $\kappa_{xx}$ 
without invoking mixed order parameter symmetry\cite{franz1}.
 
The question of the existence of the second order parameter was recently raised
once again by Sonier {\it et al}.\ in their $\mu$SR experiment 
\cite{sonier-new}. At high magnetic fields, they observed a 
flattening of $\lambda_{\rm eff}^{-2}$ (defined in these experiments
as the width of magnetic field distribution)
at low temperatures in contrast to the $T$-linear
behavior expected in a d-wave superconductor\cite{hardy,scalapino}. 
If one assumes that $\lambda_{\rm eff}^{-2}$ is proportional to the 
superfluid density $\rho_s$, then such a flattening could be indicative 
of opening of a gap in the quasiparticle excitation spectrum, which would 
result in exponentially activated behavior of $\rho_s(T)$. On the other hand,
experiment finds strong dependence of $\lambda_{\rm eff}(T\to 0)$ on the 
magnetic field. This argues against the gap
since within conventional BCS type models
opening of a gap should not affect the value of $\rho_s(T\to 0)$.
In this letter we argue that the simple relation  
$\lambda_{\rm eff}^{-2} \propto \rho_s$ is not valid for the 
penetration depth extracted in these experiments at finite fields and that 
with proper definition 
the observed behavior of $\lambda_{\rm eff}^{-2}(T)$
can be explained very naturally by a nonlocal London model
for a d-wave superconductor in which the superfluid density 
$\rho_s$ remains linear in temperature.   

High $T_c$ materials are extremely type II superconductors in the 
sense that their
coherence length $\xi$ is much smaller than their penetration depth
$\lambda$. Thus naively one would expect that the local London model 
well describes their magnetic behavior.
However, in order to study the electromagnetic response of 
a d-wave superconductor it is necessary to define a 
momentum dependent coherence length, 
$\xi_{\hat p}= v_F/\pi\Delta_{\hat p}$, 
which diverges along the node directions. This divergence gives rise to  
nonlocal dependence between the supercurrent and the vector potential.
Kosztin and Leggett \cite{kosztin} showed that in the Meissner state, 
this nonlocal relation can produce a $T^2$-dependence of the 
penetration depth, instead of linear $T$-dependence, below some crossover
temperature given by $T_{KL}^* = \Delta_0 (\xi_0 / \lambda_0)$
where $\Delta_0$ is the maximum gap, $\xi_0=v_F/\pi\Delta_0$ and 
$\lambda_0$ is the London penetration depth. 
Such an effect is field independent and can only
occur at very low temperatures. Therefore, it cannot directly explain the
$\mu$SR observation assuming that $\lambda_{\rm eff}$ is the same as
the Meissner penetration depth. We will show below that an analogous 
calculation with a new definition for the
penetration depth which is similar to its definition in $\mu$SR experiments
gives $\lambda_{\rm eff} \sim T^3$ for $T<T^*$. Here $T^*= \Delta_0 (\xi_0/d)$,
and $d \propto \sqrt{\Phi_o/B}$ is the distance between vortices, with 
$\Phi_0$ being the flux quantum. 
$T^*$ is now field dependent and could be much larger than $T_{KL}^*$.

\begin{figure}[t]
\epsfysize 6cm
\epsfbox[10 300 600 730]{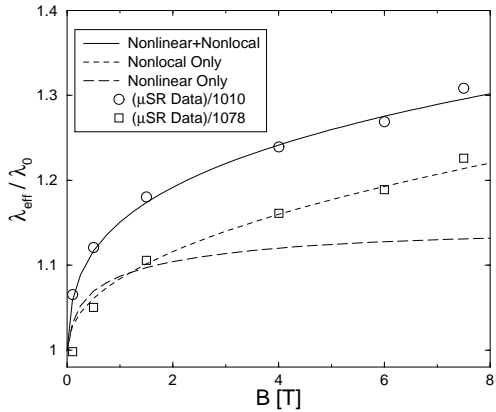}
\caption[]{Comparison between 
our zero $T$ nonlinear-nonlocal London theory (with $\lambda_0=1078$ \AA) 
and the recent $\mu$SR data\cite{sonier-new} extrapolated to $T=0$.}
\label{f1}
\end{figure}

We have studied \cite{affleck,franz2,amin}
the effect of nonlocality and nonlinearity
due to the field induced excitations at the gap nodes \cite{yip}, 
in the vortex state of a d-wave superconductor using a generalized London 
model. In Ref.\ \cite{amin} we defined an effective
penetration depth $\lambda_{\rm eff}$ so as to closely 
correspond to the quantity measured  in $\mu$SR experiments.
Fig. \ref{f1} presents the result of nonlinear-nonlocal calculation
of $\lambda_{\rm eff}$ based on the theory developed in Refs. 
\cite{franz2,amin}.
The effect of the nonlinear corrections on $\lambda_{\rm eff}$, as shown 
in Fig. \ref{f1}, 
saturates at high fields and stays effectively field independent for $B>1T$.
Most of the field dependence of $\lambda_{\rm eff}$, especially at high
fields, therefore comes from the nonlocal contributions. The effect of 
the nonlinear corrections is just to shift the value of $\lambda_{\rm eff}$ 
by a constant, which can be compensated by rescaling $\lambda_0$.
In Fig. \ref{f1}, we compare our theory with the $\mu$SR data reported in Ref.
\cite{sonier-new}. We find that it is possible to 
fit the experimental data to both nonlinear-nonlocal and nonlocal-only
curves, with fairly good agreement, by just changing the scale. 
At high fields both curves provide good fits. Even  below 1T
the agreement between the experimental data and the nonlocal-only curve 
is fairly good; although
including nonlinear corrections enhances the agreement especially at 0.1T.
The excellent agreement between our theory and the experimental data 
suggests that the same effects might be important for the temperature 
dependence of $\lambda_{\rm eff}$, and might be in fact responsible
for the $\mu$SR observation. 

The effective penetration depth in the vortex state has been also 
studied numerically by Wang and MacDonald\cite{wang1} within a lattice
model of a d-wave superconductor. This approach is capable of correctly 
accounting for the vortex core physics, which is beyond the scope of our 
London model. On the flip side, because of the system size limitations, 
the approach of Ref. \cite{wang1} is limited to relatively high fields
$B\gtrsim 10$T. The present approach is well suited to address
the low-field regime (0.1T$\lesssim B \lesssim 10$T) 
which is of greatest experimental interest in cuprates.

We now present a brief discussion of analytical and numerical 
calculations leading to the temperature and field dependence 
of $\lambda_{\rm eff}$. In order to avoid unnecessary complications we
only explicitly consider the nonlocal effects. Nonlinear corrections are 
included only by rescaling the value of $\lambda_0$, as discussed above.
A more detailed 
description of the theory is given in Refs.\ \cite{franz2,amin}. 
To linear order, the relation between the supercurrent ${\bf j}$ and 
the vector potential ${\bf A}$ in a superconductor can be 
written in Fourier space as 
%
\begin{equation}
{\bf j_{ k}}=-(c/4\pi){\bf \hat Q(k) A_{k}}.
\label{n1}
\end{equation}
where ${\bf \hat Q(k)}$ is the electromagnetic response tensor. 
Applying the linear response treatment of Gorkov equations 
generalized for an anisotropic gap, one can calculate the kernel 
${\bf \hat Q(k)}$ in weak coupling limit.
One finds \cite{franz2}
\begin{equation}
Q_{ij}({\bf k})={4\pi T\over \lambda_0^2}\sum_{n>0}\left\langle
{\Delta_{\hat p}^2\hat v_{Fi}\hat v_{Fj} \over
\sqrt{\omega_n^2+\Delta_{\hat p}^2}
(\omega_n^2+\Delta_{\hat p}^2+\gamma_{\bf k}^2)}\right\rangle,
\label{Q}
\end{equation}
where $\gamma_{\bf k}={\bf v}_F\cdot{\bf k}/2$, $\lambda_0=\sqrt{c^2/4\pi
e^2v_F^2N_F}$ is the London penetration depth, $\omega_n=\pi T(2n-1)$
are the Matsubara frequencies and the angular bracket means Fermi
surface averaging. Eq.\ (\ref{Q}) is valid for an arbitrary Fermi surface 
and gap function.  

Besides the temperature dependence contained in the Matsubara frequencies in
(\ref{Q}), the gap itself has temperature dependence which becomes important
at temperatures approaching $T_c$.
Assuming an isotropic Fermi surface, we use the simple form
$
\Delta_{\hat p}(T) = \Delta(T) \cos 2\theta,
$
for the gap where $\theta$ is the angle between the internal 
momentum ${\bf p}$ and the
x-direction. In weak coupling limit and in the absence of
magnetic field, the function $\Delta(T)$ can be obtained by solving 
the ordinary BCS gap equation which in d-wave case is
\be
{1 \over V}= T \sum_{{\bf p},\omega_n} 
{\cos^2 2\theta \over \omega_n^2 + \epsilon_p^2 + \Delta(T)^2 \cos^2 2\theta},
\label{gap}
\ee
where $V$ is the interaction potential in the d-wave channel and $\epsilon_p$
is the dispersion relation. At finite magnetic fields this equation will
change and a full self consistent calculation will be necessary to obtain 
$\Delta(T)$. However, since the magnetic fields of our interest are far
below the upper critical field $H_{c2}$, and since we focus only on
the regions outside the vortex cores, we can assume (to a good approximation) 
that Eq.\ (\ref{gap}) holds even in the presence of a weak
magnetic field. 

\begin{figure}[t]
\epsfysize 6cm
\epsfbox[10 190 560 620]{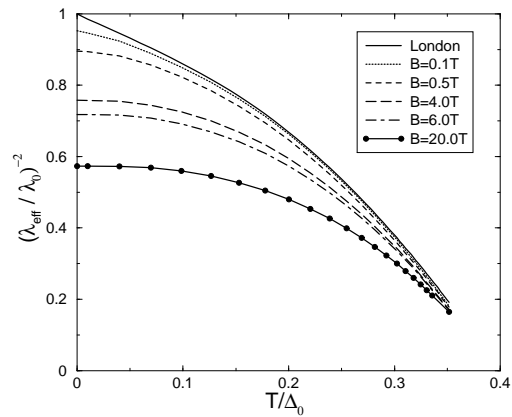}
\caption[]{Temperature dependence of $\lambda_{\rm eff}^{-2}/\lambda_0^{-2}$ 
at different magnetic fields. Here $\Delta_0 = \Delta(T=0)$.}
\label{f2}
\end{figure}

One can write the London equation purely in terms of magnetic field, by
eliminating ${\bf j}$ from  (\ref{n1}) and using 
Amp\`{e}re's law ${\bf j}=(c/4\pi)\nabla\times{\bf B}$
\begin{equation}
{\bf B_k}-{\bf k}\times[{\bf \hat Q^{-1}(k)}({\bf  k}\times{\bf B_k})]=0.
\label{london}
\end{equation}
In the vortex state it is furthermore necessary to insert a source term 
$F({\bf k})$ on the right-hand side of the Eq.\ (\ref{london}) which 
accounts for the phase winding around the vortex cores. 
We use a source function with the usual 
Gaussian cutoff $F({\bf k})=e^{-\xi_0^2 k^2/2}$ \cite{cutoff}.

As in Ref. \cite{amin}, we define $\lambda_{\rm eff}$ by
\be
\lambda_{\rm eff}^{-4} = \lambda_0^{-4} \left({\overline{\delta B^2} \over
\overline{\delta B_0^2}}\right),
\label{lmd}
\ee
where $\overline{\delta B^2}$ is
the second moment of the field distribution in the vortex lattice and 
$\overline{\delta B_0^2}$ is the same quantity for the magnetic
field $B_0(\vec r)$ obtained by solving the ordinary
London model on a triangular lattice with the same  
$\bar B$ and $\lambda_0$. 
We emphasize that this way of defining $\lambda_{\rm eff}$
is roughly equivalent to the way it is computed from the $\mu$SR data. 
Using (\ref{london}) and (\ref{lmd}) we get
\be
\lambda_{\rm eff}^{-4}
= C \sum_{{\bf k} \ne 0} {e^{-\xi_0^2 k^2} \over (1+ {\cal L}_{ij} k_i k_j)^2}
\approx C \sum_{{\bf k} \ne 0}{e^{-\xi_0^2 k^2} \over({\cal L}_{ij} k_i k_j)^2},
\label{lmdeff}
\ee
where ${\bf k}$ are the reciprocal lattice wave vectors and
\be 
{\cal L}_{ij}({\bf k})={Q_{ij}({\bf k}) \over {\rm det}\ \hat{\bf Q}
({\bf k})} \quad, \quad 
C^{-1}=\sum_{{\bf k} \ne 0} {e^{-\xi_0^2 k^2} \over k^4}. 
\ee
Notice that only wave-vectors of O($d^{-1})$ and larger contribute to the
vortex lattice field distribution and the response to a
magnetic field at these wave-vectors can be much different
than at zero wave-vector in a superconductor with gap nodes.
At low $T$, only the regions near the gap nodes are important in the calculation
of $Q_{ij}({\bf k})$. We can therefore linearize the gap as $\Delta_{\hat{p}}
\approx \Delta_\theta \equiv 2 \Delta(T) \theta$ where $\theta$ is the angle 
measured from 
the node, and let ${\bf v}_F$ in Eq. (\ref{Q}) only take the node
directions. As a result, $Q_{ij}$ will be diagonal in the $45^\circ$ rotated
frame. 

Defining $q_{1,2}=(k_x \pm k_y)/\sqrt{2}$ we can write $Q_{ij}=\lambda_0^{-2}
[1+K(q_i,T)]\delta_{ij}$. We then
have
\ba
&& {\cal L}_{11}({\bf q}) = \lambda_0^{2}[1- K(q_2,T)] \nn \\
&& {\cal L}_{22}({\bf q}) = \lambda_0^{2}[1- K(q_1,T)] \\
&& {\cal L}_{12}({\bf q}) = {\cal L}_{21}({\bf q}) = 0, \nn
\ea
This is actually a Taylor expansion in $K$ which is small at low
$B$ and $T$. Substituting into (\ref{lmdeff}) we get
\be
{\lambda_{\rm eff} \over \lambda_0} 
=1 - {\tilde{C} \over 2} \sum_{\tilde{\bf q} \ne 0} 
{\tilde{K}(\tilde{q}_1,T)\tilde{q}_2^2 +  
\tilde{K}(\tilde{q}_2,T)\tilde{q}_1^2
\over \tilde{q}^6},
\label{dlmd}
\ee
where $\tilde{\bf q}={\bf q}d$, $\tilde{C}=Cd^4$ and 
$\tilde{K}(x,T)=K(x/d,T)$.
We have also taken the upper cutoff  $\zeta_d^{-1}=d/\xi_0$ 
to infinity since it does not affect our calculations (we keep $\zeta_d$
finite in our numerical calculation but our results are insensitive to 
its exact value). 

At $T=0$ for $\tilde{q} \zeta_d \ll 1$ one can write \cite{kosztin}
$\tilde{K}(\tilde{q},0) \approx - (\pi^2 / 8) 
\zeta_d \ \tilde{q}$. 
Substituting back into (\ref{dlmd}) we get 
\be
\delta \lambda_{\rm eff}(T=0)\ \propto \ \zeta_d \ \sim \ \sqrt{\bar B}.
\ee
This is in complete agreement with our numerical calculation 
and $\mu$SR data \cite{sonier-new} (c.f. Fig. \ref{f1}), and
also with other calculations \cite{vekhter2}.

At $T>0$, Eq. (\ref{dlmd}) can be written as
\ba
{\delta \lambda_{\rm eff} \over \lambda_0} 
= - {\tilde{C}\over 2} \sum_{\tilde{\bf q} \ne 0} 
{\delta \tilde{K}(\tilde{q}_1,T)\tilde{q}_2^2 + \delta 
\tilde{K}(\tilde{q}_2,T)\tilde{q}_1^2
\over \tilde{q}^6},
\ea
where $\delta\lambda_{\rm eff}=\lambda(B,T)-\lambda(B,0)$ and 
$\delta \tilde{K}(\tilde{q},T) = \tilde{K}(\tilde{q},T)- 
\tilde{K}(\tilde{q},0)$.
We follow Ref. \cite{kosztin} by writing
\be
\delta \tilde{K}(\tilde{q},T)=\tilde{K}(0,T) F\left({\tilde{q}
\over t} \right),
\label{dK}
\ee
where $t=T/T^*$, $\tilde{K}(0,T) \approx -2(\ln 2)\ T/\Delta_0$ is the kernel 
obtained in local approximation,  and $F(z)$ is a universal function which
can be approximated by
\ba
F(z) &\approx & 1-c_1z \qquad {\rm for} \quad z<2 \nn \\
F(z) &\approx & c_0/z^2 \qquad \ \ {\rm for} \quad z>2.
\label{F}
\ea
Here, $c_0$ and $c_1$ are constants.
Substituting (\ref{F}) and (\ref{dK}) into (\ref{dlmd}) we find that for
$t \ll 1$, $F(z)$ falls into $z>2$ regime for all reciprocal lattice
vectors $\tilde{\bf q}$. 
This immediately gives $\delta \lambda_{\rm eff} \propto T^3$. At higher
temperatures on the other hand, there will exist significant
number of points with $z<2$ which 
would give linear $T$ behavior. Thus in general

\ba
&& \delta \lambda_{\rm eff} \propto T^3
\qquad {\rm for} \quad \ \ \  T \ll T^* \nn \\
&& \delta \lambda_{\rm eff} \propto T
\qquad \ {\rm for} \quad T^* \ll T \ll T_c.
\ea

\begin{figure}[t]
\epsfysize 6cm
\epsfbox[10 190 560 620]{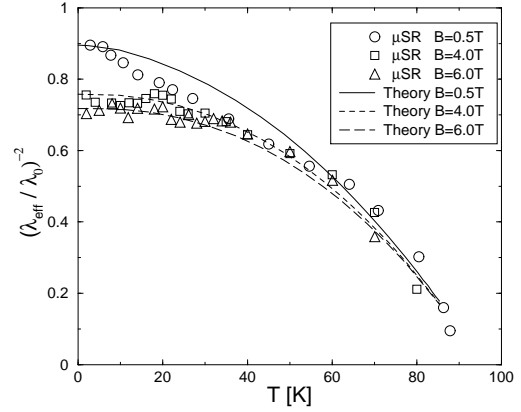}
\caption[]{A comparison between our nonlocal calculation and $\mu$SR data.}
\label{f3}
\end{figure}

Fig.\ \ref{f2} shows the results of our numerical calculation of
$\lambda_{\rm eff}$ as a function of temperature.  
We take $\lambda_0=1078$\AA, which is the value that produces the best fit
to the experimental data (cf. Fig. 3), and $\kappa = \lambda_0/\xi_0
=68$.  We also assume a triangular vortex
lattice aligned with $x$ and $y$ ($a$ and $b$) directions. 
We have tried changing the core size as well as the
shape of the vortex lattice and our results remained unchanged.
The penetration depth is no longer linear at low temperatures; unlike 
the superfluid density (the upper curve). 
The deviation from linearity is stronger at higher
fields in complete agreement with the $\mu$SR observation\cite{sonier-new}
and previous theoretical work\cite{wang1}. At high enough fields, the $T$
dependence is in complete agreement with the $T^3$-form obtained analytically.
The curves for different fields join to a single curve at higher temperatures. 
This is essentially because the nonlocal
corrections are most pronounced for the quasiparticles close to the
gap nodes. At higher temperatures the response becomes dominated by the
quasiparticles far from the nodes which feel much weaker nonlocal effects.

Fig.\ \ref{f3} presents a 
comparison between our results and the experimental data. In order to 
get a good fit, we set $\Delta_0 \equiv \Delta(0)=2.65 T_c$ which 
is also what one obtains from the gap equation (\ref{gap}). 
We have normalized both experimental and theoretical data with
 $\lambda_0=1078$ \AA . The agreement between our theory and the data 
is good at $B=4$T and 6T, but at $B=0.5$T the theory shows less 
linearity at low temperatures compared to the experimental
data,  although they agree fairly well at $T=0$ and high $T$.
 
As we mentioned earlier, a complete calculation should include nonlinear
corrections \cite{amin,yip} in addition to the nonlocal corrections 
considered here. However, as apparent from Fig.\ \ref{f1}, 
at $T=0$ the main source of the field dependence of $\lambda_{\rm eff}$ is 
the nonlocal effect. Since this field dependence is closely related to the 
flattening of the curves at low $T$, it is reasonable to assume that 
nonlocal effects also dominate the $T$-dependence. The main discrepancy 
between the theory and experiment in Fig.\ \ref{f3} is the lack of linearity 
at 0.5T; however it is difficult to envision how nonlinear corrections could 
cure this. 
Given the inherent uncertainties in the extraction of $\lambda$ from the
$\mu$SR data we consider the overall agreement to be reasonable even 
without inclusion of the nonlinear effects.

In summary, we have calculated the field and temperature dependence of 
the effective penetration depth 
$\lambda_{\rm eff}$ from a nonlocal London model of a d-wave superconductor. 
We used a definition of $\lambda_{\rm eff}$ which  permits a direct 
comparison to the $\mu$SR experimental data. Our results exhibit 
a $T^3$-dependence in the $\lambda_{\rm eff}^{-2}$ curve 
below $T^*=\Delta_0 (\xi_0/d) \sim \sqrt{B}$,
quantitatively consistent with the experimental data on YBCO\cite{sonier-new}. 
This flattening
has nothing to do with the reduction of the superfluid density or 
opening a true gap at high magnetic fields. Rather, it is a consequence
of the nonlocal response of a d-wave superconductor which modifies the
magnetic field distribution in the vortex lattice as compared to an ordinary 
London model. Thus, as pointed out previously\cite{amin,wang1}, it is
essential to make a distinction between the London penetration
depth $\lambda_L$ (as measured e.g. in the microwave experiment \cite{hardy})
and the effective penetration depth  $\lambda_{\rm eff}$ deduced from the
magnetic field profile in a $\mu$SR experiment. In the present
model $\lambda_{\rm eff}$ flattens at low temperatures and finite field while
$\lambda_L$ remains linear in $T$. 

The authors are indebted to J. E. Sonier for supplying the data and frequent 
discussions and also to I. Herbut for stimulating conversations. 
The work was supported in part by  NSERC, the CIAR, and NSF
grants DMR-9415549 (M.F.) and PHY-94-07194 (I.A.).

\end{document}